\documentclass[11pt,fleqn,twoside]{article}
\usepackage{amsfonts,latexsym}
\makeatletter
\newcommand{\prava}{\footnotesize\it
\begin{flushright}
\begin{minipage}{18cm}
Copyright \copyright 1998 by B. Lu, Y. He and G. Ni
\end{minipage}
\end{flushright}}

\newcommand{\name}[1]{\begin{flushleft}
                       \LARGE \bf #1
                       \end{flushleft}\vspace{-3mm}}

\newcommand{\Author}[1]{\begin{flushleft}
                       \it #1 \end{flushleft}}

\newcommand{\Adress}[1]{\begin{flushleft}
                       \it #1 \end{flushleft}}

\newcommand{\Date}[1]{\begin{flushleft}
                      \small  \it #1 \end{flushleft}}

\newcommand{\ehkol}{Author \ name}
\newcommand{\ohkol}{Article \ name}
\renewcommand{\@evenhead}{
\hspace*{-3pt}\raisebox{-15pt}[\headheight][0pt]{\vbox{\hbox to \textwidth
{\thepage \hfil \ehkol}\vskip4pt \hrule}}}
\renewcommand{\@oddhead}{
\hspace*{-3pt}\raisebox{-15pt}[\headheight][0pt]{\vbox{\hbox to \textwidth
{\ohkol \hfil \thepage}\vskip4pt\hrule}}}
\renewcommand{\@evenfoot}{}
\renewcommand{\@oddfoot}{}

\newcommand{\be}{\begin{equation}}
\newcommand{\ee}{\end{equation}}
\newcommand{\ba}{\hspace*{-5pt}\begin{array}}
\newcommand{\ea}{\end{array}}
\newcommand{\p}{\partial}
\newcommand{\ds}{\displaystyle}

\makeatother

\begin{document}
\setcounter{page}{140}

\renewcommand{\theequation}{\thesection.\arabic{equation}}

\thispagestyle{empty}

\renewcommand{\ehkol}{B. Lu, Y. He and G. Ni}
\renewcommand{\ohkol}{A Method for Obtaining Darboux Transformations}

\begin{flushleft}
\footnotesize \sf
Journal of Nonlinear Mathematical Physics \qquad 1998, V.5, N~2,
\pageref{lu-fp}--\pageref{lu-lp}.
\hfill
{{\sc Letter}}
\end{flushleft}

\vspace{-5mm}

{\renewcommand{\footnoterule}{}
{\renewcommand{\thefootnote}{}  \footnote{\prava}}

\name{A Method for Obtaining Darboux Transformations}\label{lu-fp}

\Author{Baoqun LU~$^{\dag\star}$, Yong HE~$^{\ddag}$ and  Guangjiong NI
$^{\dag}$}

\Adress{$^{\dag}$~Department of Physics, Fudan University, 200433,
Shanghai, P.R. China\\[1mm]
$^{\star}$~Institute of Mathematics, Fudan University, 200433,
Shanghai, P.R. China\\
~(mailing address)\\[1mm]
$^{\ddag}$~Department of Applied Mathematics, the Central South
University \\
~of Technology of China, 410083, Changsha, P.R. China
}

\Date{Received December 12, 1997; Accepted February 27, 1998}

\begin{abstract}
\noindent
In this paper we give a method to obtain Darboux transformations
(DTs) of integrable equations. As an example we give a DT of the
dispersive water wave equation.  Using the Miura map, we also obtain
the DT of the Jaulent-Miodek equation.
\end{abstract}

\section{Introduction}
For integrable equations which can be solved by the Inverse Scattering
Transform, there exist B\"acklund transformations (BTs)~[1]. These
transformations were f\/irst discovered for the Sine-Gordon equation at
the end of the 19th century. Usually they are treated as nonlinear
superpositions, which allow one to create new
  solutions of a nonlinear evolution equation from a f\/inite number of
  known solutions. In practice, BTs are not very
  straightforward to apply in the construction of multisolutions. On
the other hand,
the Darboux transformation (DT) is a very convenient way of
constructing new solutions of nonlinear integrable
equations~[2]; the algorithm is purely algebraic and can be continued
successively.  Therefore, it is interesting to transform BTs  into DTs.

Many  integrable equations   of the form
\be
u_t=K(u)
\ee
possess the recursion operator $\Phi$ with the property  called hereditary
symmetry [3, 4, 5, 6], and they possess a Lax pair
\be
\ba{l}
\Phi\sigma=\lambda\sigma,\\
\sigma_t=K_u\sigma.
\ea
\ee
Here $K_u$ is the Fr\'echet derivative of $K$ with respect to $u$.
Two interesting questions are raised: ``How is the DT related
to the Lax pair (1.2)?'' and ``What happens to the symmetry
$\sigma$ under a BT?''

 In this paper, we will study the above problems. Section 2 gives the general
method to obtain DTs of integrable equations by using symmetry. As an
example, Section 3 gives the DT of the dispersive water equation. In
Section 4, we obtain
  the DT of the Jaulent-Miodek equation by using the Miura map.
  These DTs are not easily obtained by other well-known methods.

\section{The Method}

\setcounter{equation}{0}

Suppose  that the equation (1.1) has a BT of the form
\be
 B(u,u[1])=0.
\ee
Now, we suppose that
\be
 u_{u}[1][\sigma_1]=0,
\ee
which means that the symmetry $\sigma_1$ is transformed into $0$
under the BT,
where $\sigma_1$ is the eigenfunction of (1.2) with $\lambda=\lambda_1$.

Then taking Fr\'echet derivative of $B=0$, we have
\be
B_u[\sigma_1]= B_u[\sigma_1]+B_{u[1]}[u_{u}[1]\sigma_1]=0.
\label{lu:eq:B}
\ee

Fuchssteiner and Aiyer have showed that the KdV equation, the Burgers
equation, and the CDGSK equation admit this relation [7, 8, 9].

This formula gives the transformation relation between $u,\sigma_1$
and $u[1]$:
\be
 u[1]=F(u,\sigma_1).
\label{lu:eq:u}
\ee
At this point we can directly check whether (\ref{lu:eq:u}) is a
BT.
If so, we can conclude that (\ref{lu:eq:B}) is true,
and we also have the transformation for eigenfunctions
\be
 \sigma[1]=u_u[1][\sigma]=F_u[\sigma]+F_{\sigma_1}[\sigma_{1u}\sigma].
 \label{lu:eq:ss}
 \ee
Relations (\ref{lu:eq:u}) and (\ref{lu:eq:ss}) is called the DT
of (1.1).

\medskip

\noindent
{\bf Remark 2.1.} Here we give a method to calculate $\sigma_{1u}[\cdot]$.
Because  $\Phi\sigma_1=\lambda\sigma_1$, we have
\[
 \sigma_{1u}[\cdot]=-(\Phi-\lambda_1)^{-1}\Phi_u[\cdot]\sigma_1.
\]
We can apply the factorization method to calculate
 $(\Phi-\lambda_1)^{-1}$.

\medskip

\noindent
{\bf Remark 2.2.} When (\ref{lu:eq:u}) and (\ref{lu:eq:ss})
 is the DT
\[
\Phi(u[1])\sigma[1]=\lambda\sigma[1],
\]
and (\ref{lu:eq:ss}) is the symmetry of $u_{t}[1]=K(u[1])$, the result
[10] shows that
(\ref{lu:eq:u}) and (\ref{lu:eq:ss}) is
a DT for the hierarchy $u_t=\Phi^n K(u)$.

\medskip

\noindent
{\bf Remark 2.3.} Relation (\ref{lu:eq:u})  reveals the connection
among the BT,
symmetry, and strong symmetry operator. We conjecture that
(\ref{lu:eq:u}) may be right for all equations
which possess strong symmetry operators.
 For some $(1+1)$-dimension equations
($u_t=K(u)$), their usual Lax pair
\[
L\phi=\lambda\phi,\qquad \phi_t=A\phi
\]
can be transformated into
\[
\Psi\sigma=\mu\sigma,\qquad \sigma_t=K_u\sigma
\]
by a transformation $\sigma=f(\phi)$.  We can then obtain the DT with
respect to the usual Lax pair.

\medskip


\section{The DT of the Dispersive Water Wave Equation}

\setcounter{equation}{0}

In this section we study the dispersive water wave equation (DWW) [11, 12]
\be
v_t=K(v),
\label{lu:eq:DW}
\ee
where
 \[
\ba{l}
  v=(q,r)^T,\\[2mm]
\ds   K(v)=\left(\frac{1}{2}(2qr-q_x)_x, \frac{1}{2}(r_x+r^2+2q)_x\right)^T
  \ea
\]
and $T$ denotes the transpose of vectors. The DWW equation was
studied systematically by Kupershmidt~[11].

System (\ref{lu:eq:DW}) has the following Lax representation~[11].
\be
L\phi=\lambda\phi, \label{lu:eq:L1}
\ee
\be
\phi_t=\frac{1}{2}(L^2)_{+}\phi,
\label{lu:eq:L2}
 \ee
in which
\[
L=D+(D-r)^{-1}\circ q, \qquad D=\frac{\partial}{\partial x},
\qquad D^{-1}\circ D=D\circ D^{-1}=1
\]
and $(\cdot )_{+}$ is the projection to the purely dif\/f\/erential part
of the operator, $(L^2)_{+}=D^2+2q.$
Here we denote the operator $A$ acting on the operator $B$ by $A\circ B$,
and the operator $A$ action on a function  $f$ by $Af$.

The system (\ref{lu:eq:DW}) possesses a strong symmetry operator
\be
\Phi(v)=\left(\ba{cc}
 -D+r & 2q+q_x\circ D^{-1}\\[1mm]
2 & D+D\circ r\circ D^{-1}
\ea\right).
\ee

So (\ref{lu:eq:DW}) is the following integrable condition
\be
\Phi\sigma=\lambda\sigma,
 \label{lu:eq:c1}
\ee
\be
\sigma_t=K_v \sigma,
 \label{lu:eq:c2}
\ee
where $\sigma=(\sigma_1,\sigma_2)^T$.

It is dif\/f\/icult to get a DT of (\ref{lu:eq:DW}) with respect to
(\ref{lu:eq:L1}), (\ref{lu:eq:L2}). In fact, we did not f\/ind
any DT for (\ref{lu:eq:L1}), (\ref{lu:eq:L2}) until now.

Let us turn  to (\ref{lu:eq:c1}), (\ref{lu:eq:c2}).

\medskip

\noindent
{\bf Theorem 3.1.} {\it Let $\sigma_{,1}=(\sigma_{1,1},\sigma_{2,1})^T$ denote
the solution of (\ref{lu:eq:c1}), (\ref{lu:eq:c2}) with
$\lambda=\lambda_1$. We then have the DT
\be
q[1]=q-\left(\frac{D^{-1}\sigma_{1,1}}{D^{-1}\sigma_{2,1}}\right)_x,
\label{lu:eq:q[1]}
\ee
\be
r[1]=r+\left(\ln\frac{D^{-1}\sigma_{1,1}+\sigma_{2,1}}{D^{-1}\sigma_{2,1}}
\right)_x,
\label{lu:eq:r[1]}
\ee
\be
\sigma_1[1]=\sigma_1-\left(\frac{B}{D^{-1}\sigma_{2,1}}\right)_x,
\label{lu:eq:s1}
\ee
\be
\sigma_2[1]=\sigma_2-\left(\frac{B+\sigma_{2}D^{-1}\sigma_{2,1}}{D^{-1}
\sigma_{1,1}+\sigma_{2.1}}\right)_{x},
 \label{lu:eq:s2}
\ee
where
\[
B=D^{-1}(\sigma_1 D^{-1}\sigma_{2,1}+\sigma_2 D^{-1}\sigma_{1,1} ),
\]
that is, (\ref{lu:eq:q[1]}) and (\ref{lu:eq:r[1]}) is a new solution of
(\ref{lu:eq:DW}). Moreover, (\ref{lu:eq:q[1]}), (\ref{lu:eq:r[1]})
    and (\ref{lu:eq:s1}), (\ref{lu:eq:s2}) satisfy (\ref{lu:eq:c1}),
     (\ref{lu:eq:c2}). Furthermore, this is the DT of the hierarchy
$v_t=\Phi(v)^nK(v)$.}

\medskip

\noindent
{\bf Proof.} (i) From practice, we f\/irst suppose
$q[1]=q+(ln\phi_1)_{xx} $ is a part of the BT.
Substituting $\phi_1=e^{D^{-2}(q[1]-q)}$ into (\ref{lu:eq:L1}) with
$\lambda=\lambda_1$, we f\/ind
\be
(D^{-1}(q[1]-q))^{2}+q[1] + \lambda_1 r-(r+\lambda_1)D^{-1}(q[1]-q)=0.
\ee
  From (\ref{lu:eq:B}) we f\/ind
\[
-(2D^{-1}\sigma_{1,1}+\sigma_{2,1})D^{-1}(q[1]-q)+
(r+\lambda_1)D^{-1}\sigma_{1,1} +\lambda_1\sigma_{2,1}=0.
\]
On the other hand, (\ref{lu:eq:c1}) gives
\[
2D^{-1}\sigma_{1,1} +\sigma_{2,1}=(\lambda_1-r)D^{-1}\sigma_{2,1}.
\]
These two identities imply (\ref{lu:eq:q[1]}).

Suppose $(q[1],r[1])$ satisfy (\ref{lu:eq:DW}), then
\be
2(q[1]-q)_t=(2q[1]r[1]-2qr-(q[1]-q)_x)_x ,
\ee
\be
-2\left(\frac{D^{-1}\sigma_{1,1}}{D^{-1}\sigma_{2,1}}\right)_t
=2q[1]r[1]-2qr +\left(\frac{D^{-1}\sigma_{1,1}}{D^{-1}\sigma_{2,1}}
\right)_{xx}.
\ee
Using (\ref{lu:eq:c2}), we have
\be
\ba{l}
\ds   2q[1]r[1]-2qr+\left(\frac{D^{-1}\sigma_{1,1}}{D^{-1}\sigma_{2,1}}
\right)_{xx}=-\frac{1}{D^{-1}\sigma_{2,1}}
               (2\sigma_{1,1} r+2q\sigma_{2,1}-\sigma_{1,1x})\\[3mm]
\ds \phantom{2q[1]r[1]-2qr+\left(\frac{D^{-1}\sigma_{1,1}}{D^{-1}\sigma_{2,1}}
\right)_{xx}=}
+\frac{D^{-1}\sigma_{1,1}}{(D^{-1}\sigma_{2,1})^2}(\sigma_{2,1x}+2r\sigma_{2,1}
               +2\sigma_{1,1}).
 \label{lu:eq:x}
 \ea
 \ee

We note that (\ref{lu:eq:c1}) yields
\be
r=\lambda_1-\frac{\sigma_{2,1}+2D^{-1}\sigma_{1,1}}{D^{-1}\sigma_{2,1}},
\label{lu:eq:r}
\ee
\be
q=\frac{1}{(D^{-1}\sigma_{2,1})^2}(\sigma_{1,1}D^{-1}\sigma_{2,1}+
(D^{-1}\sigma_{1,1})^2).
\label{lu:eq:q}
\ee
Substituting the above two identities and (\ref{lu:eq:q[1]}) into
(\ref{lu:eq:x}),  we obtain, after some calculations, (\ref{lu:eq:r[1]}).

We can easily prove that (\ref{lu:eq:q[1]}) and (\ref{lu:eq:r[1]}) satisfy
 (\ref{lu:eq:c1}) and (\ref{lu:eq:c2}), so (\ref{lu:eq:q[1]}),
 (\ref{lu:eq:r[1]}) is a BT of (\ref{lu:eq:DW}).

(ii) From (\ref{lu:eq:r}) and (\ref{lu:eq:q})
\[
 v_{\sigma_1}D= \frac{1}{D^{-1}\sigma_{2,1}}\left(
 \ba{cc}
 \ds D+\frac{2D^{-1}\sigma_{1,1}}{D^{-1}\sigma_{2,1}}
 & \ds -\frac{\sigma_{1,1}}{D^{-1}\sigma_{1,1}}
 -2\left(\frac{D^{-1}\sigma_{1,1}}{D^{-1}\sigma_{2,1}}\right)^2\\[4mm]
-2 & \ds -D+\frac{\sigma_{2,1}+2D^{-1}\sigma_{1,1}}{D^{-1}\sigma_{2,1}}
\ea
 \right).
\]
Now, we solve the equation
\[
v_{\sigma}D\left(\ba{c} a_1\\ a_2\ea
\right)=
\left(\ba{c} \sigma_{1}\\ \sigma_{2} \ea\right)
\]
that is,
\be
a_{1x}+2\frac{D^{-1}\sigma_{1,1}}{D^{-1}\sigma_{2,1}}a_{1}-
\left(\frac{\sigma_{1,1}}{D^{-1}\sigma_{2,1}}
   +2\left(\frac{D^{-1}\sigma_{1,1}}{D^{-1}\sigma_{2,1}}
   \right)^2\right)a_2=\sigma_{1}D^{-1}\sigma_{2,1},
   \label{lu:eq:a1}
\ee
\be
-2a_1-a_{2x}+\left(\frac{\sigma_{2,1}+2D^{-1}\sigma_{1,1}}{D^{-1}
\sigma_{2,1}}\right)a_2=   \sigma_2 D^{-1}\sigma_{2,1}.
   \label{lu:eq:b1}
\ee
$\ds (\ref{lu:eq:a1})+(\ref{lu:eq:b1})\frac{D^{-1}\sigma_{1,1}}
{D^{-1}\sigma_{2,1}}$,  we f\/ind
\[
 a_1 - a_2 \frac{D^{-1}\sigma_{1,1}}{D^{-1}\sigma_{2,1}}=B
 \]
with
\[
B=D^{-1}((D^{-1}\sigma_{2,1}) \sigma_1 +\sigma_2 D^{-1}\sigma_{1,1} ).
\]
Hence
\[
\ba{l}
\ds a_1=B-(D^{-1}\sigma_{1,1})
\left(2D^{-1}\frac{B}{D^{-1}\sigma_{2,1}}+D^{-1}\sigma_2 \right),\\[4mm]
\ds a_2 =-2D^{-1}\sigma_{2,1}D^{-1}\frac{B}{D^{-1}\sigma_{2,1}}
-D^{-1}\sigma_2 D^{-1}\sigma_{2,1}.
 \ea
 \]

Now we can calculate $\sigma[1]$ from (\ref{lu:eq:q[1]}),
(\ref{lu:eq:r[1]}):
\[
\ba{l}
\ds \sigma[1]=v_v [1] \sigma  \\[4mm]
\ds \phantom{\sigma[1]}=\sigma+D\left(\ba{cc}
\ds   \frac{-1}{D^{-1}\sigma_{2,1}}
    & \ds \frac{D^{-1}\sigma_{1,1}}{(D^{-1}\sigma_{2,1})^2}\\[3mm]
\ds \frac{1}{D^{-1}\sigma_{1,1}+\sigma_{2,1}}
    &\ds \frac{D}{D^{-1}\sigma_{1,1} +\sigma_{2,1}}-\frac{1}{D^{-1}\sigma_{2,1}}
    \ea\right)
    D^{-1}(v_{\sigma_1})^{-1} \sigma\\[5mm]
\ds \phantom{\sigma[1]}
=\sigma+D\left(\ba{cc}
\ds -\frac{1}{D^{-1}\sigma_{2,1}}
    & \ds \frac{D^{-1}\sigma_{1,1}}{(D^{-1}\sigma_{2,1})^2}\\[3mm]
\ds    \frac{1}{D^{-1}\sigma_{1,1}+\sigma_{2,1}}
    &\ds \frac{D}{D^{-1}\sigma_{1,1} +\sigma_{2,1}}-
    \frac{1}{D^{-1}\sigma_{2,1}}\ea\right)
     \left(\ba{c} a_1 \\ a_2\ea\right)\\[5mm]
\ds \phantom{\sigma[1]} =
\sigma-D\left(\ba{c}\ds  \frac{B}{D^{-1}\sigma_{2,1}}\\[3mm]
\ds  \frac{B+\sigma_2D^{-1}\sigma_{2,1}}
 {\sigma_{2,1} +D^{-1}\sigma_{1,1} }
\ea\right).
\ea
\]
This completes the proof.

\medskip

\noindent
{\bf Remark 3.1.}
Let $w=\sigma_x$, then (\ref{lu:eq:c1}),
(\ref{lu:eq:c2}) can be written in a more simple
form:
\[
\ba{l}
 -w_{xx}+rw_{1x}+2qw_{2x}+q_x w_2 =\lambda w_{1x},\\[1mm]
 2w_{1x}+(w_{2x}+rw_2)_x =\lambda w_{2x},\\[1mm]
2w_{1t}=2rw_{1x}+2w_{2x}q-w_{1xx},\\[1mm]
2w_{2t}=2w_{2xx}+2rw_{2x}+2w_{1x}.
\ea
\]

The DT given in Theorem 3.1 becomes
\[
\ba{l}
\ds q[1]=q-\left(\frac{w_{1,1}}{w_{2,1}}\right)_x, \\[3mm]
\ds r[1]=r+\left(\ln\frac{w_{1,1}+w_{2,1x}}{w_{2,1}}\right)_x, \\[3mm]
\ds w[1]=w-\left(\ba{c}\ds  \frac{B}{w_{2,1}} \\[3mm]
\ds  \frac{B+w_{2,1x}(1)w_{2,1}}{w_{1,1} + w_{2,1x}}
\ea\right),
\ea
\]
with
\[
 B=D^{-1}(w_{2,1} w_{1x}+w_{2x}w_{1,1} ).
 \]

\section{The DT of the Jaulent-Miodek equation}
\setcounter{equation}{0}

The Jaulent-Miodek equation takes the form [13, 14]
\be
 u_t=H(u)=\Psi(u)u_x,
  \label{lu:eq:JM}
 \ee
where
\[
u=\left(\ba{c} u_0\\u_1\ea\right),
\qquad
 \Psi(u)=\left(\ba{cc} 0 & \ds \left(\frac{1}{4}D^3 +
 \frac{1}{2}u_{0}\circ D  +\frac{1}{2}D\circ u_0\right)\circ D^{-1}\\[3mm]
   1 & \ds \left(\frac{1}{2}u_1\circ D+
   \frac{1}{2}D\circ u_1 \right)\circ D^{-1}\ea
   \right)
\]
and $\Psi$ is a strong symmetry operator.
So (\ref{lu:eq:JM}) possesses the Lax pair
\be
\Psi(u)\psi=\lambda\psi,
 \label{lu:eq:j1}
 \ee
 \be
 \psi_t=H_u\psi.
 \label{lu:eq:j2}
 \ee
 Usually (\ref{lu:eq:JM}) is derived from the following
 spectral problem [12, 13, 14]:
 \be
 L\phi=\phi_{xx}+(u_0+\lambda u_1)\phi=\lambda^2\phi,
 \ee
where the time evolution of the wave function $\phi$ has the form
\be
\phi_t=P\phi=\left(\frac{1}{2}p\circ D-\frac{1}{4}p_x\right)\phi,
\ee
with $p=2+\lambda u_1$. Then
\[
 L_t - [P,L]=p_x L
 \]
gives rise to (\ref{lu:eq:JM}).  The BT of
 (\ref{lu:eq:JM})  was given by Tu [13].
 It is not easy to apply this BT to construct
 new solutions.

An invertible Miura map [12]
\be
q=u_0 +\frac{1}{4} (u_1)^2-\frac{1}{2}u_{1x},
 \label{lu:eq:m1}
\ee
\be
r=u_1
 \label{lu:eq:m2}
 \ee
brings (\ref{lu:eq:JM}) into the DWW (\ref{lu:eq:DW}).

 The Miura map (\ref{lu:eq:m1}), (\ref{lu:eq:m2}) gives the
 relation of the eigenfunctions
 between (\ref{lu:eq:j1}), (\ref{lu:eq:j2}) and (\ref{lu:eq:c1}),
 (\ref{lu:eq:c2}):
 \be
  \psi=u_v \sigma=\left(\ba{cc} 1 & \ds \frac{D}{2}-\frac{r}{2}\\[2mm]
  0  & 1\ea\right)\sigma
  \ee
  \be
  \phantom{\psi}
=\left(\ba{c} \ds \sigma_1+\frac{\sigma_{2x}}{2}-\frac{r}{2}\sigma_2\\[2mm]
 \sigma_2
  \ea\right).
\ee
Therefore,
\[
D^{-1}\sigma_1=D^{-1}\psi_1-\frac{\psi_2}{2}+\frac{D^{-1}(u_1\psi_2)}{2}
=\left(\lambda-\frac{u_1}{2}\right)D^{-1}\psi_2-\frac{\psi_2}{2},
\]
\be
u_1[1]  =u_1+\left(\ln\frac{D^{-1}\sigma_{1,1}+\sigma_{2,1}}
{D^{-1}\sigma_{2,1}}\right)_x=u_1+E,
     \label{lu:eq:u1}
     \ee
where
\[
E=\left(\ln\left(\lambda_1-\frac{u_1}{2}-\frac{\psi_{2,1}}{2D^{-1}\psi_{2,1}}
\right)\right)_x.
\]
\be
\ba{l}
\ds u_0[1]-u_0=(q[1]-q)+\frac{1}{2}(r[1]-r)_x -
\frac{1}{4}(r[1]-r)(r[1]+r)\\[3mm]
\ds \phantom{u_0[1]-u_0}=
\left(\frac{1}{2}u_1+\frac{\psi_{2,1}}{2D^{-1}\psi_{2,1}}\right)_x+
\frac{1}{2}E_x          -\frac{1}{4}E(E+2u_1),
\label{lu:eq:u0}
\ea
\ee
\[
\ba{l}
\ds B=D^{-1}(\sigma_1 D^{-1}\sigma_{2,1}+\sigma_2 D^{-1}\sigma_{1,1} )\\[3mm]
\ds \phantom{B}=(D^{-1}\psi_{2,1})
\left(\left(\lambda_1-\frac{u_1}{2}\right)D^{-1}\psi_2
 -\frac{\psi_2 }{2}\right)\\[3mm]
\ds \phantom{B} -D^{-1}\left(\psi_{2,1} \left(\lambda_1 -
\frac{u_1}{2}\right)D^{-1}\psi_2 -\psi_2 \left(\lambda_1
 -\frac{u_1}{2}\right)D^{-1}\psi_{2,1}\right),
\ea
\]
\be
\psi_2[1]=\psi_2-F,
\label{lu:eq:p2}
\ee
\be
\psi_1[1]=\psi_1-\left(\frac{B}{D^{-1}\psi_{2,1}}\right)_x-
\frac{F_x}{2}-\frac{1}{2}(E\psi_{2,1}- F(u_1+E)),
         \label{lu:eq:p1}
\ee
with
\[
 F=D\left(\frac{B+\psi_2 D^{-1}\psi_{2,1}}{\left(\lambda_1-
 \frac{1}{2}u_1\right)D^{-1}\psi_{2,1}
+\frac{1}{2}\psi_{2,1}}\right).
\]

\noindent
{\bf Theorem 4.1.}
{\it Suppose $(u,\psi_{,1})$ satisfies (\ref{lu:eq:j1}), (\ref{lu:eq:j2})
with $\lambda=\lambda_1$, then the transformation defined by
(\ref{lu:eq:u1}), (\ref{lu:eq:u0}), (\ref{lu:eq:p2}), (\ref{lu:eq:p1})
 is a DT of (\ref{lu:eq:j1}), (\ref{lu:eq:j2}).}

\section{Conclusion }
In this paper, we have presented a method to obtain  DTs of integrable
 equations. This method  can be apply to the DWW equation, the KdV equation,
 a shallow water equation [15]  and other integrable equations.
 We think the relation (\ref{lu:eq:B}) is very important because
it  reveals the relation between BT,
symmetry, and strong symmetry of the corresponding
equation. We hope that there will be further study in this direction.

\subsection*{Acknowledgements}
This work was supported by Shanghai Science and Technology "Morning
Star" planning of China.

 \label{lu-lp}
}
\end{document}